\documentclass[preprint]{aastex}

\slugcomment{To appear in the ApJS Special Issue on the ``Chandra Carina Complex Project'', May 2011}

\shorttitle{Infrared properties of the X-ray emitting YSOs in the Carina Nebula}
\shortauthors{Preibisch et al.}

\begin{document}

\title{Near-Infrared properties of the X-ray emitting young stellar objects in the 
Carina Nebula}


\author{Thomas Preibisch}
\affil{Universit\"ats-Sternwarte M\"unchen,  Ludwig-Maximilians-Universit\"at,
              Scheinerstr.~1, 81679 M\"unchen, Germany}

\author{Simon Hodgkin, Mike Irwin, James R.~Lewis}
\affil{Cambridge Astronomical Survey Unit,
             Institute of Astronomy, Madingley Road, Cambridge, CB3 0HA, UK}

\author{Robert R.~King}
\affil{Astrophysics Group, College of Engineering, Mathematics and Physical Sciences, University of Exeter, Exeter EX4 4QL, UK}

\author{Mark~J. McCaughrean\altaffilmark{1}}
\affil{European Space Agency, Research \& Scientific Support Department, ESTEC, Postbus 299, 2200 AG Noordwijk, The Netherlands} 
\altaffiltext{1}{on leave from: Astrophysics Group, College of Engineering, Mathematics, and Physical Sciences, University of Exeter, Exeter EX4 4QL, UK}

\author{Hans Zinnecker\altaffilmark{2}}
\affil{Deutsches SOFIA Institut, Universit\"at Stuttgart, Pfaffenwaldring 31,
 70569 Stuttgart, Germany, and NASA-Ames Research Center,
MS 211-3, Moffett Field, CA 94035, USA}
\altaffiltext{2}{on leave from: Astrophysikalisches Institut Potsdam, An der 
             Sternwarte 16, 14482 Potsdam, Germany}

\and

\author{Leisa Townsley and Patrick Broos}
\affil{Department of Astronomy \& Astrophysics,
             Pennsylvania State University, University Park PA 16802, USA
}



\begin{abstract}
The Great Nebula in Carina (NGC~3372) is the best target
to study in detail the process of violent massive star formation 
and the resulting feedback effects of
cloud dispersal and triggered star formation.
While the population of massive stars is rather well studied,
the associated low-mass stellar population was largely unknown
up to now. The near-infrared study in this paper
builds on the results of 
the {\it Chandra} Carina Complex Project (CCCP), that
detected 14\,368 X-ray sources in the 1.4 square-degree survey region,
an automatic source classification study that
classified 10\,714 of these X-ray sources as very likely
young stars in Carina, and an analysis of the clustering properties of
the X-ray selected Carina members.
In order to determine physical properties of the X-ray selected stars,
most of which were previously unstudied,
we used HAWK-I at the ESO VLT
to conduct a very deep near-IR survey with sub-arcsecond angular
resolution, covering  an
area of about 1280 square-arcminutes.
The HAWK-I images 
reveal more than 600\,000 individual infrared sources, whereby 
objects as faint as
$J\approx 23$, $H\approx 22$, and $K_{\rm s}\approx 21$
are detected at S/N $\ge 3$. 
While less than half of the {\it Chandra} X-ray sources have
counterparts in the 2MASS catalog, the $\sim 5$~mag deeper
HAWK-I data reveal infrared counterparts to
6636 (= 88.8\%) of the 7472 {\it Chandra} X-ray sources in the HAWK-I field.
We analyze near-infrared color-color and color-magnitude diagrams
to derive information about the extinctions, infrared excesses (as
tracers for circumstellar disks),
ages, and masses of the X-ray selected objects.
The near-infrared properties agree well with the results of the 
automatic X-ray source classification, showing that the remaining
contamination in the X-ray selected sample of Carina members
is very low ($\lesssim 7\%$).
The shape of the $K$-band luminosity function of the X-ray selected Carina members
agrees well with that derived for the Orion Nebula Cluster, 
suggesting that, down to the X-ray detection limit around $0.5-1\,M_\odot$,
the shape of the IMF in Carina is consistent with that in Orion 
(and thus the field IMF).
The fraction of stars with near-infrared excesses 
is rather small, $\la 10\%$, but shows
considerable variations between individual parts of the complex.
The distribution of extinctions for the diskless stars 
ranges from
$\sim 1.6$~mag to $\sim 6.2$~mag (central 80th percentile), 
clearly showing a
considerable range of differential extinction between individual stars
in the complex. 
\end{abstract}




\section{Introduction}


The Great Nebula in Carina \citep[NGC 3372; see][for an overview]{SB08}
is the best galactic analog of giant extragalactic 
H\,II and starburst regions and a superb location to study the 
physics of violent massive star formation and the resulting feedback effects, 
including cloud dispersal and triggered star formation. 
At a distance of 2.3\,kpc \citep{Smith02}, it represents the nearest 
southern region with a large massive stellar population \citep[65 
O-type stars; see][]{Smith06}. Among these are several
of the most massive and luminous stars known in our Galaxy, 
most notably the
famous Luminous Blue Variable $\eta$\,Car. 
Most of these very massive
stars reside in the clusters Tr16, Tr~14, and Tr~15, for 
which ages between $<$1 and several Myr have been estimated.

In the past, the 
Carina Nebula was usually considered to be just
an evolved H{\sc II} region, devoid of active star formation.
However, new sensitive observations have changed this view drastically 
during recent years.
The region contains more than $10^5\,M_\odot$ of gas and dust
 \citep[see][]{Yonekura05,SB07,Preibisch_Laboca},
and deep infrared observations 
showed clear evidence of ongoing star formation in these clouds.
Several very young stellar objects
\citep{Megeath96,Mottram07} and a spectacular young embedded cluster
\citep[the ``Treasure Chest Cluster''; see][]{Smith05} have been 
found in the molecular clouds, 
 a deep HST H$\alpha$ imaging survey revealed
dozens of jet-driving young stellar objects \citep{Smith10a}, and
\textit{Spitzer} surveys located numerous embedded protostars  throughout the
Carina complex \citep{Smith10b,Povich11a}. The formation of these young stars
was probably triggered by the
advancing ionization front, and thus a substantial population
of very young, partly embedded stars seems to be present.


While the unobscured  population of high-mass stars $(M \geq 20\,M_\odot)$
in the Carina Nebula Complex (= CNC hereafter)
is rather well known and characterized \citep{Smith06}, 
the much fainter low-mass $(M \leq 2\,M_\odot)$ stellar 
population remained largely unexplored until now.
However,
a good knowledge of the low-mass stellar content is essential
for any determination of the global properties of this region, 
which are the key towards understanding the  star formation process
in the CNC. 
Important topics,  for which the low-mass
stellar population plays a crucial role, include
the  initial mass function (IMF) \citep[e.g.,][]{Bastian10} 
of the region,
the star formation history \citep[e.g.,][]{Briceno07}, 
and the effects of the feedback
from the numerous massive stars on the formation and evolution 
of low-mass stars and their protoplanetary disks 
\citep[e.g.,][]{Whitworth04,Throop05}.

%

\smallskip


The obvious first step of any study of the low-mass stellar
population is the
identification of the individual low-mass stars. However,
this is quite a difficult task in the case of the CNC
for several reasons.
First, due to its location very close to the galactic 
plane ($b \approx -0.6\degr$) 
and near the tangent point of a spiral arm, any optical
and infrared observations of the CNC suffer from 
extremely strong field star contamination and confusion problems. 
Second, these problems are
amplified by the strongly variable and highly position dependent
pattern of cloud extinction across the CNC, that ranges
from $A_V \approx 1$~mag at some positions up to $A_V \approx 10$~mag
only a few arcminutes away \citep{Rowles09}.
The dark clouds in the CNC thus do {\em not} serve 
as an opaque\footnote{Since the maximum extinction of 
$A_V \approx 10$~mag corresponds to only $A_K \approx 1$~mag
in the near-infrared, 
the clouds are mostly transparent for wavelengths $\gtrsim 2\,\mu$m.} 
``screen'' 
that would hide the background sources, 
but rather introduce a complicated pattern of strong variations 
in the extinction of the individual complex members as well as
the background objects seen through these clouds.
Third, strong and complex nebular emission (in particular H$\alpha$ in the
optical and Br$\gamma$ in the $K$-band) is present from interstellar gas
ionized by the CNC OB stars.
Furthermore, due to the relatively large distance, the low-mass stars are
relatively faint, making an individual spectroscopic identification
of young low-mass stars \citep[e.g.~by their Lithium lines or
gravity-sensitive lines; see, e.g.,][]{Preibisch02,Slesnick08}
unfeasible.

Until recently, essentially all of the few known
young low-mass stars in the CNC
were identified by infrared excess emission \citep[e.g.,][]{Ascenso07}, 
which is a tracer of
circumstellar material surrounding young stellar objects. 
However, there are (at least) two problems with excess-selected 
samples:
First, from studies of other regions it is well known that
infrared excess emission in young stars 
disappears on timescales of just a few Myr \citep[e.g.,][]{Briceno07};
at an age of $\sim 3$~Myr 
only $\sim 50\%$ of the young stars still 
show near-infrared excesses, and by $\sim 5$~Myr this is reduced to $\sim 15\%$.
Since the expected ages of most young stars in the CNC are
a few Myr, {\em any excess-selected sample will be 
highly incomplete}.
On the other hand, excess-selected samples 
can be {\em strongly contaminated}
by background sources, since  
evolved Be stars, carbon stars,
planetary nebulae, star-forming galaxies, and AGN 
can show near-infrared excesses very
similar to those of young stars \citep[e.g.,][]{Mentuch09,Oliveira09,Rebull09}.
This background contamination is particularly
strong in Carina due to its position at the galactic plane and
the moderate level of cloud extinction.
These two factors strongly limit the usefulness of a
near-infrared excess selected sample of young stars in the case of the CNC.
\medskip

Sensitive X-ray observations can provide
a very good solution of this problem, since  one can
detect the young stars by their strong X-ray emission
\citep[e.g.,][]{Feigelson07}
and efficiently
discriminate them from the numerous
older field stars in the survey area.
X-rays are equally sensitive to young stars which have already
dispersed their circumstellar disks, thus avoiding the bias
introduced when selecting
samples based only on infrared excess.
Many X-ray studies of star forming regions have demonstrated
the success of this method
\citep[see, e.g.,][]{PZ02,Broos07,coronet,Wang10}.
Also, the
 relations between the X-ray properties and basic stellar
properties in young stellar populations
are now  very well established from very deep X-ray
observations such as the 
{\it Chandra} Orion Ultradeep Project (COUP)
\citep[see][]{Feigelson05,Preibisch_coup_orig,PF05}.
\bigskip

Although the CNC has been observed with basically all 
X-ray observatories of the last few decades,
only the {\it Chandra} X-ray Observatory 
has good enough angular resolution ($< 1''$) to allow 
a proper identification of the individual X-ray sources in such a
crowded region.
The basis of the study presented in this paper are the results of a
wide-field X-ray survey performed in the 
{\it Chandra} Carina Complex Project (CCCP).
General aspects of this survey and the resulting X-ray data are described 
in full detail in \citet{Townsley11a} and \citet{Broos11a}. 
While these {\it Chandra} data provide an unbiased (although luminosity-limited)
sample of the young stars in the region, the X-ray data alone
do not yield much information about the properties of the 
individual stars.
For these purposes, {\em deep optical or near-infrared data 
are fundamentally required\/} 
in order to determine key stellar parameters including mass, age, and
circumstellar disk properties.
Numerous optical and infrared observations exist already
\citep[see, e.g.,][]{Smith87,Kaltcheva93,DeGioia01,Tapia03},
but they are either too shallow to detect the
low-mass stellar populations, or cover only relatively small
parts of the complex \citep[see, e.g.,][]{Sanchawala07,Smith04,Ascenso07}.    
This was the motivation for our new
deep wide-field near-infrared survey of the CNC discussed in this paper.

In this paper we briefly summarize basic results of the
X-ray observations (Sect.~2), describe the near-infrared observations 
(Sec.~3),
use the near-infrared data to identify infrared counterparts
of the X-ray sources (Sec.~4),
determine key properties of the X-ray selected young stellar 
populations in the CNC (Sec.~\ref{results.sec}), and
finally draw conclusions about the star formation process in the CNC
(Sect.~6).

\section{Basic results on the CNC X-ray sources from the CCCP}

The {\it Chandra} Carina Complex Project (CCCP)  has mapped
the CNC with a mosaic of 22  individual
ACIS-I pointings, each with an exposure time of $\sim 60$~ksec 
($\sim 17$ hours).
A complete overview of the project can be found in
\citet{Townsley11a}. Data analysis followed procedures described by 
\citet{Broos10}.
After extensive source detection efforts,
a final merged list of 14\,368 individual X-ray sources\footnote{Note
that \citep[as described in more detail in][]{Townsley11a} the star
 $\eta$~Car, which is a very strong X-ray source,
is {\em not} included in the CCCP source list, because it is so bright
that it caused very strong pileup in our {\it Chandra}
data. $\eta$~Car is therefore {\em not} considered in the analysis of this paper.}
was compiled \citep{Broos11a}.
As in any X-ray observation,
there must be some degree of contamination
by foreground stars as well as by background stars
and extragalactic sources \citep[see][]{Getman11}. 
Since these different kinds of contaminants have different typical
X-ray, optical, and infrared properties, 
a statistical approach 
was used to compute the probability that, 
based on its individual source properties,
a given X-ray source is a member of the CNC
or one of three different contaminant classes.
The results of this classification, which is described in full
detail in \cite{Broos11b}, are as follows:
716 X-ray sources are classified as very likely foreground stars (class = H1),
16 as very likely galactic background stars (class = H3),
877 X-ray sources as very likely extragalactic (AGN) contaminants (class = H4),
and 10\,714 X-ray sources are classified as very likely
young stars in the CNC (class = H2).
Finally, the classification remained ambiguous for a total of 
2045 X-ray sources; these  sources are denoted as ``unclassified''.
It should be noted that, due to the statistical nature of this analysis,
these classifications 
do not provide final evidence about the physical nature of the
individual X-ray sources. 
The review of these classifications by \citet{Broos11b}
suggested that among the  very weak X-ray sources classified as
likely Carina members there may still be a small degree of contamination
by foreground stars and AGN (see Sect.~\ref{contamination.sec}); 
on the other hand, some fraction of the unclassified X-ray sources 
may in fact be young stars in the CNC.

\citet{Feigelson11} studied the spatial distribution and the
clustering properties of the X-ray sources classified as
likely Carina members.
They identified 20 principal clusters of X-ray stars (most of 
which correspond to known optical clusters in the CNC),
31 small groups of X-ray stars outside the major
clusters, and a widely dispersed, but highly populous, 
distribution of more than 5000 X-ray stars.

The well established correlation between X-ray luminosities and
bolometric luminosities of young stars
\cite[see][]{Preibisch_coup_orig,Telleschi07}
leads to correlations between the near-infrared
magnitudes and the X-ray fluxes (or source count rates)
that can be used to infer the
expected NIR magnitudes for the {\it Chandra} X-ray sources detected
in the CNC.
The observed correlations
in the Orion/COUP data \citep{Preibisch_coup_orig} predict that the
weakest detected X-ray sources in the Carina mosaic should have
near-infrared magnitudes in the range
$J \sim 15.7^{+3}_{-2}$ and $K_s \sim 14.7\pm 2 $ (quoted ranges
are 90th percentiles).
Comparing these values to the 
nominal 2MASS Survey completeness limits for crowded locations
near the galactic plane, i.e.~$J\approx 14.8$ 
and $K_s \approx  13.3$ \citep[][]{Skrutskie06}, 
it follows
that a substantial fraction of the X-ray sources is expected 
to remain undetected in the 2MASS data.
This expectation is confirmed when considering the results of the
matching of X-ray source positions to the
2MASS catalog \citep{Broos11a}: only
6194 ($43.1\%$) of the 14\,368 X-ray sources have valid counterparts
in the 2MASS catalog.
Furthermore, many of these 2MASS matches have upper magnitude limits
in one or more bands; just
4502 of these 2MASS matches have valid $J$-, $H$-, and $K_s$-band photometry.
The fact that 
68.7\% of the X-ray sources have no or incomplete
NIR photometry from 2MASS 
clearly illustrates the need for a much deeper 
near-infrared survey of the CNC.


\section{HAWK-I Observations of the Carina Nebula Complex}

\subsection{Observations and data analysis}

\begin{figure}[h] \epsscale{0.5}
\plotone{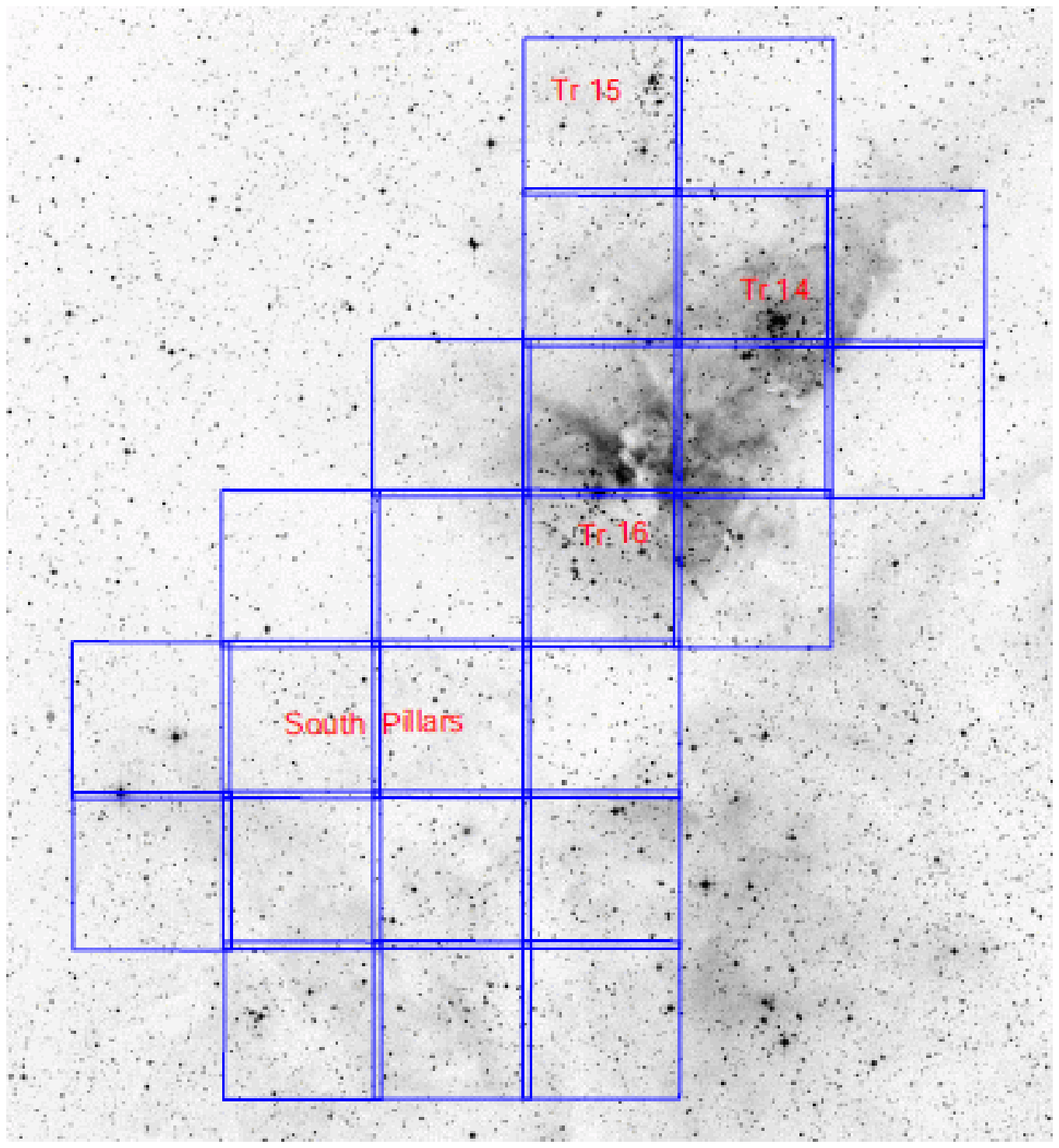}
\figcaption{Outline of the HAWK-I mosaic pattern shown on the
    red DSS image of the Carina Nebula displayed in negative gray scale.
The individual mosaic fields are shown as boxes with a size
of $7.5' \times 7.5'$. The most important regions in the CNC
are indicated.  North is up and east to the left.
\label{dss-hawki.fig}}
\end{figure}

HAWK-I \citep[see][]{HAWKI08} is the new near-infrared imager at the 
ESO 8\,m Very Large Telescope, available since the year 2008. 
It is equipped with a
mosaic of four Hawaii 2RG $2048\times 2048$ pixel detectors with
a scale of $0.106''$ per pixel.
The camera
has a field of view on the sky of $7.5' \times 7.5'$
with a small cross-shaped gap of $\sim 15''$ between the four
detectors. 
The observations of the CNC were performed 
from 24~to 31~January 2008 in service mode
as part of the scientific
verification program for HAWK-I.
The observing conditions were generally good, with a
typical seeing of $0.5'' - 0.8''$ (FWHM).
The data discussed in this paper are derived from images taken through
the standard
broadband $J$, $H$, and $K_{\rm s}$ filters.
Our survey consists of a
mosaic of 24 contiguous HAWK-I fields (see Fig.~\ref{dss-hawki.fig})
covering a total area of about 1280 square-arcminutes; it 
includes the central part of the Nebula with $\eta$~Car and Tr~16, 
the clusters Tr~14 and Tr~15, and large 
parts of the South Pillars region. A full description of the
HAWK-I data will be given in \citet{Preibisch_HAWKI}.

\begin{figure}[h] \epsscale{0.8}
\plotone{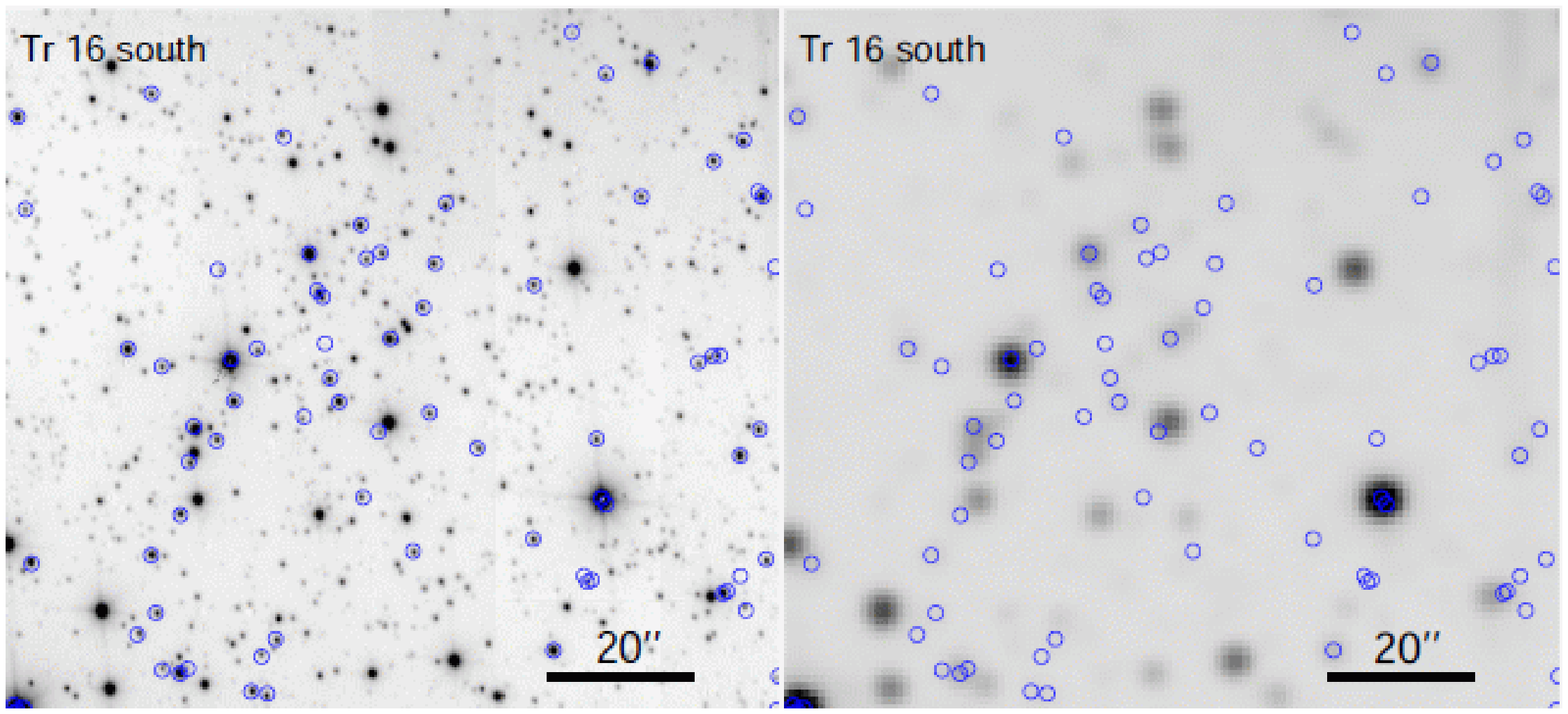}
\vspace{1mm}

\plotone{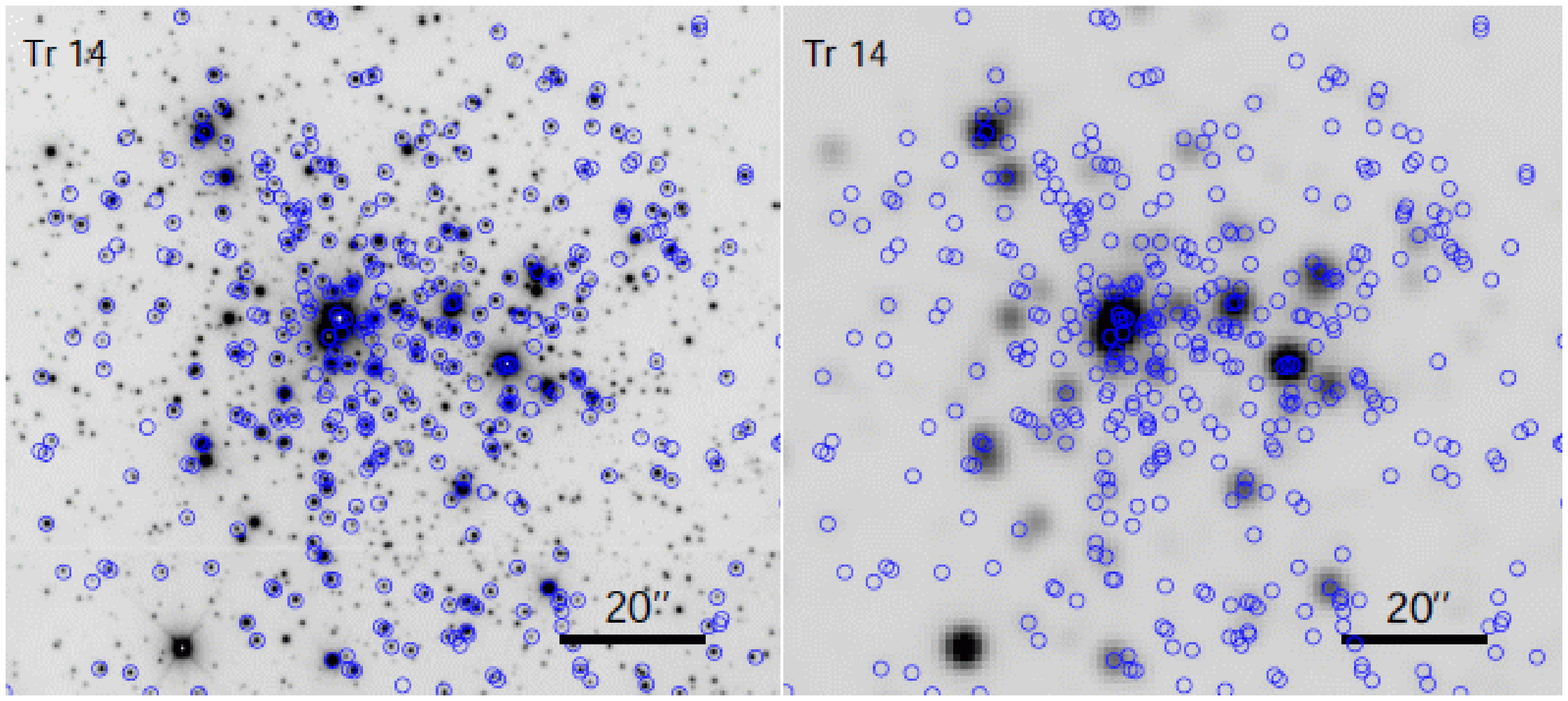}
\vspace{1mm}

\plotone{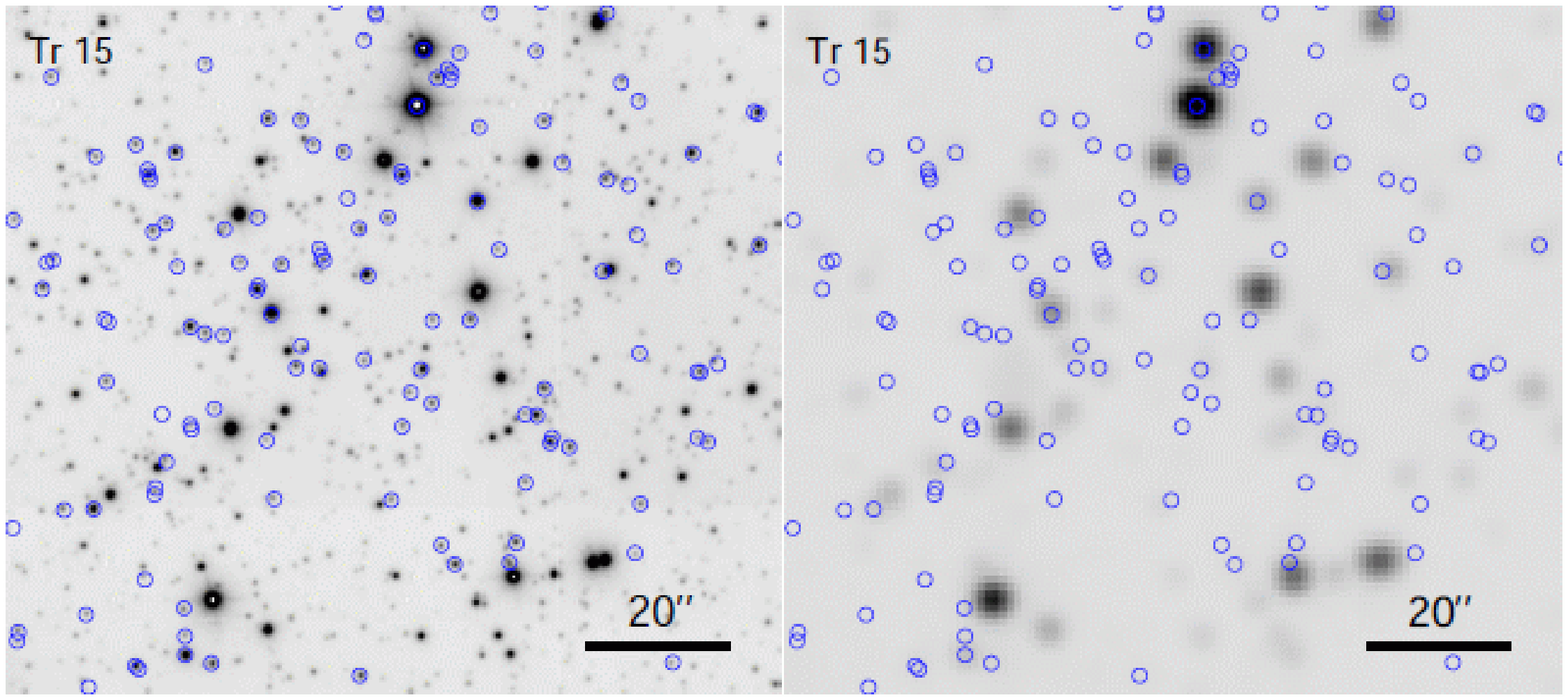}
\figcaption{Comparison of HAWK-I (left) and 2MASS (right) $H$-band
images of three selected regions, covering a field in the southern
part of Tr~16 (centered at the position
$\alpha(J2000)=10^{\rm h}\,45^{\rm m}\,10^{\rm s}$, $\delta(J2000)=-59\degr\,42'\,46''$), the cluster Tr~14, and
the southern part of the cluster Tr~15. X-ray source positions
are marked by circles.
\label{hawki-2mass.fig}}
\end{figure}

The aim of the HAWK-I survey was to
detect the full stellar population
through extinctions of $A_V = 15$~mag
{\em simultaneously} in the $J$-, $H$-,
{\em and} $K_{\rm s}$-band.
With  magnitude limits of
$J=22.4$, $H=20.7$, and $K_{\rm s}=19.7$,
all stars with  masses down to $0.1\,M_\odot$ and an age of 3~Myr
should be detectable 
according to the pre-main-sequence models of \citet{Baraffe98}.
The required  exposure times for S/N=10 detections
are 12, 8, and 5 minutes in $J$, $H$, and $
K_{\rm s}$.
We used a 5-point dither pattern with $40''$ steps
and took co-added images with detector integration times of 3 seconds
at each mosaic position.
As examples,  we show in Fig.~\ref{hawki-2mass.fig} HAWK-I
images of three selected regions of the mosaic and compare them
to the corresponding 2MASS images.

All HAWK-I data were processed and calibrated by the Cambridge
Astronomical Survey Unit using pipeline software originally
designed for the analysis of the UKIRT Infrared Deep Sky Survey
\citep[see][]{Irwin04}.
After the initial standard reduction steps 
for each individual image, the dithered images of each mosaic
position were stacked together, thereby ``removing'' the
gaps between the four infrared arrays.
After the automatic detection of point-like sources in the images
\citep[see][]{Irwin85}
the source extraction software  measured
aperture photometry for each detected source.
The photometric analysis and calibration followed closely the
procedures described in detail in \citet{Hodgkin09}. 
The photometric calibrators are drawn from stars in the 
2MASS Point Source Catalog, which are
present in large
numbers (several hundred) in each HAWK-I mosaic field.

\begin{figure} \epsscale{0.5}
\plotone{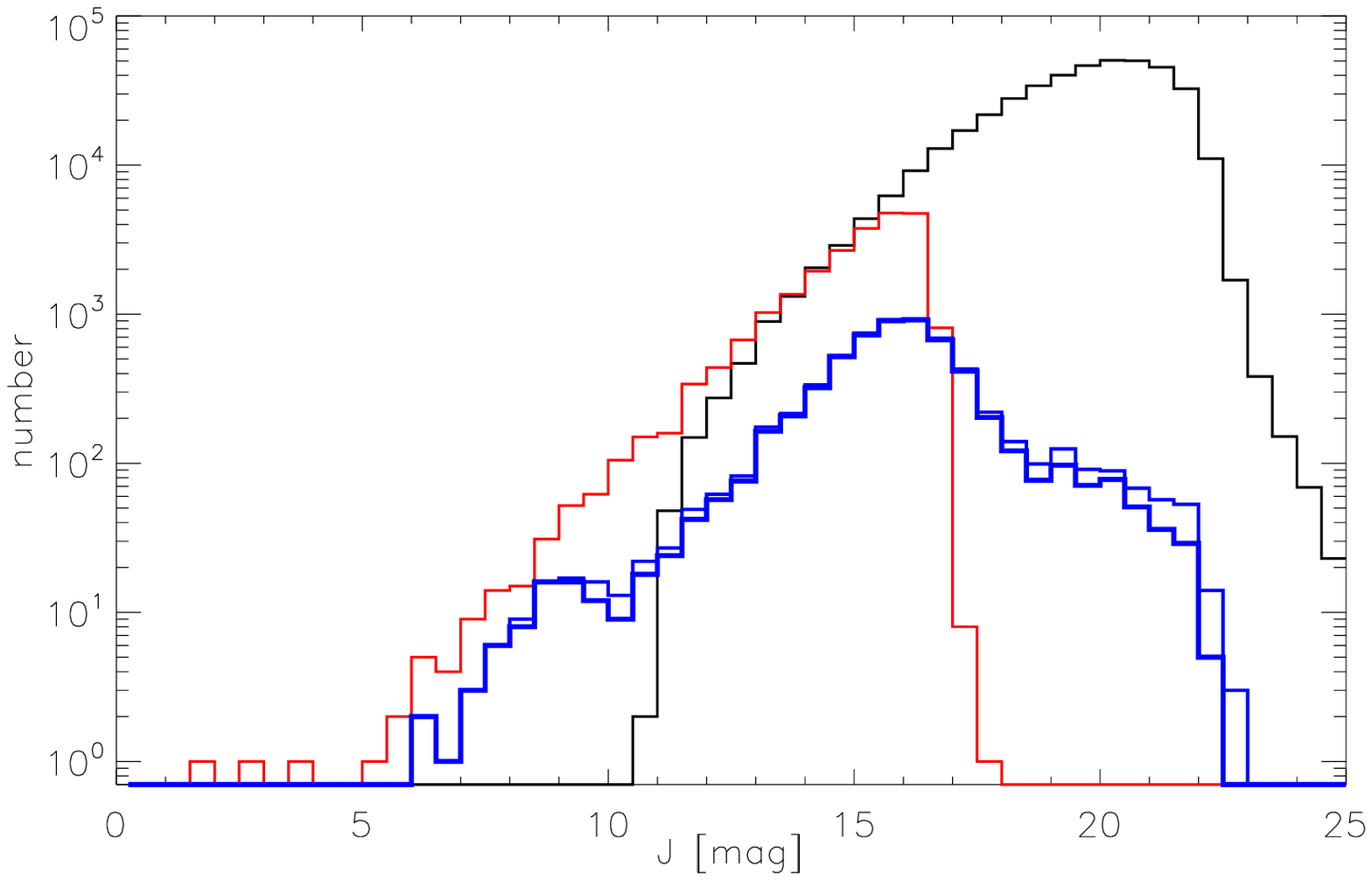}
\plotone{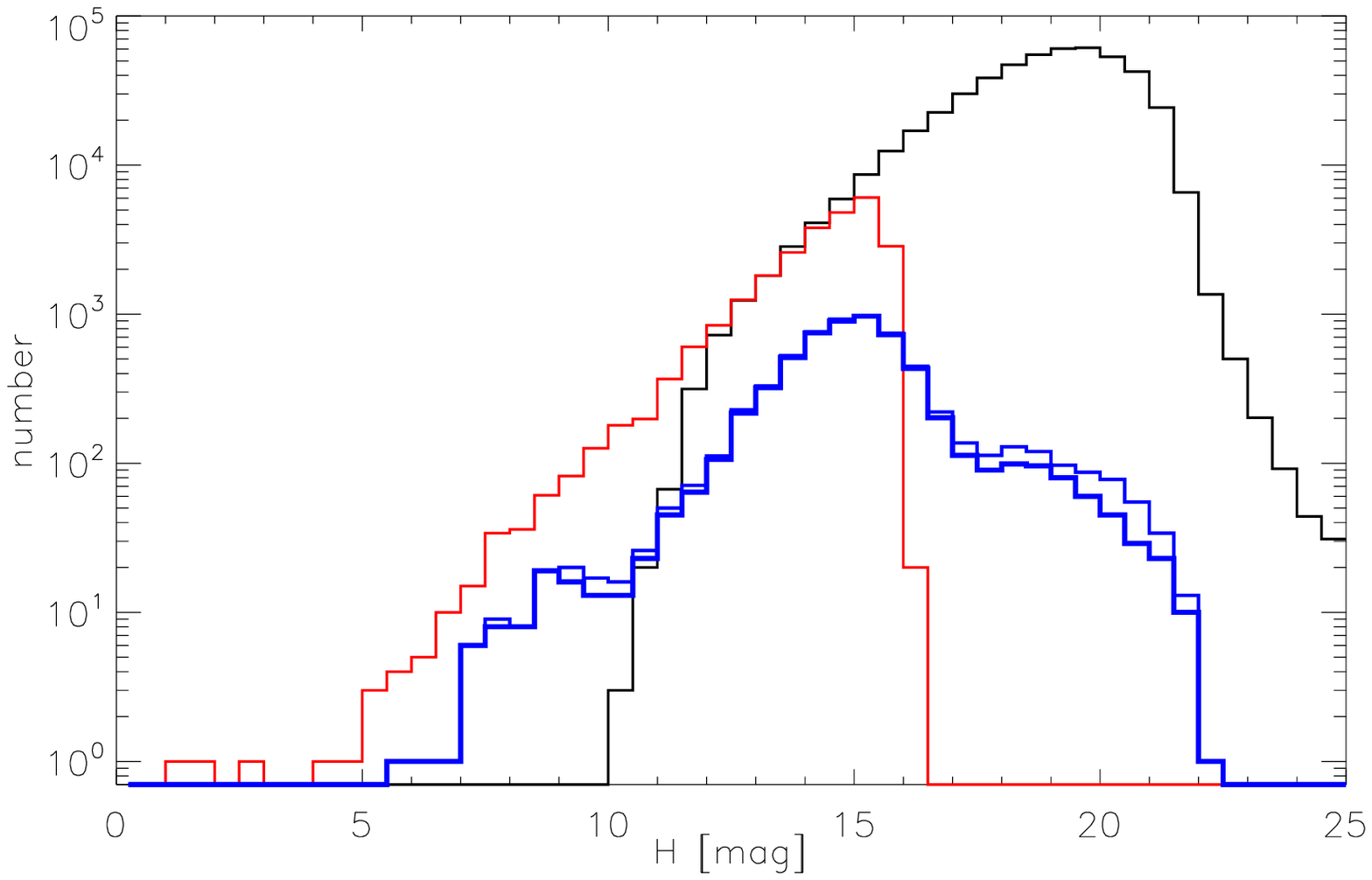}
\plotone{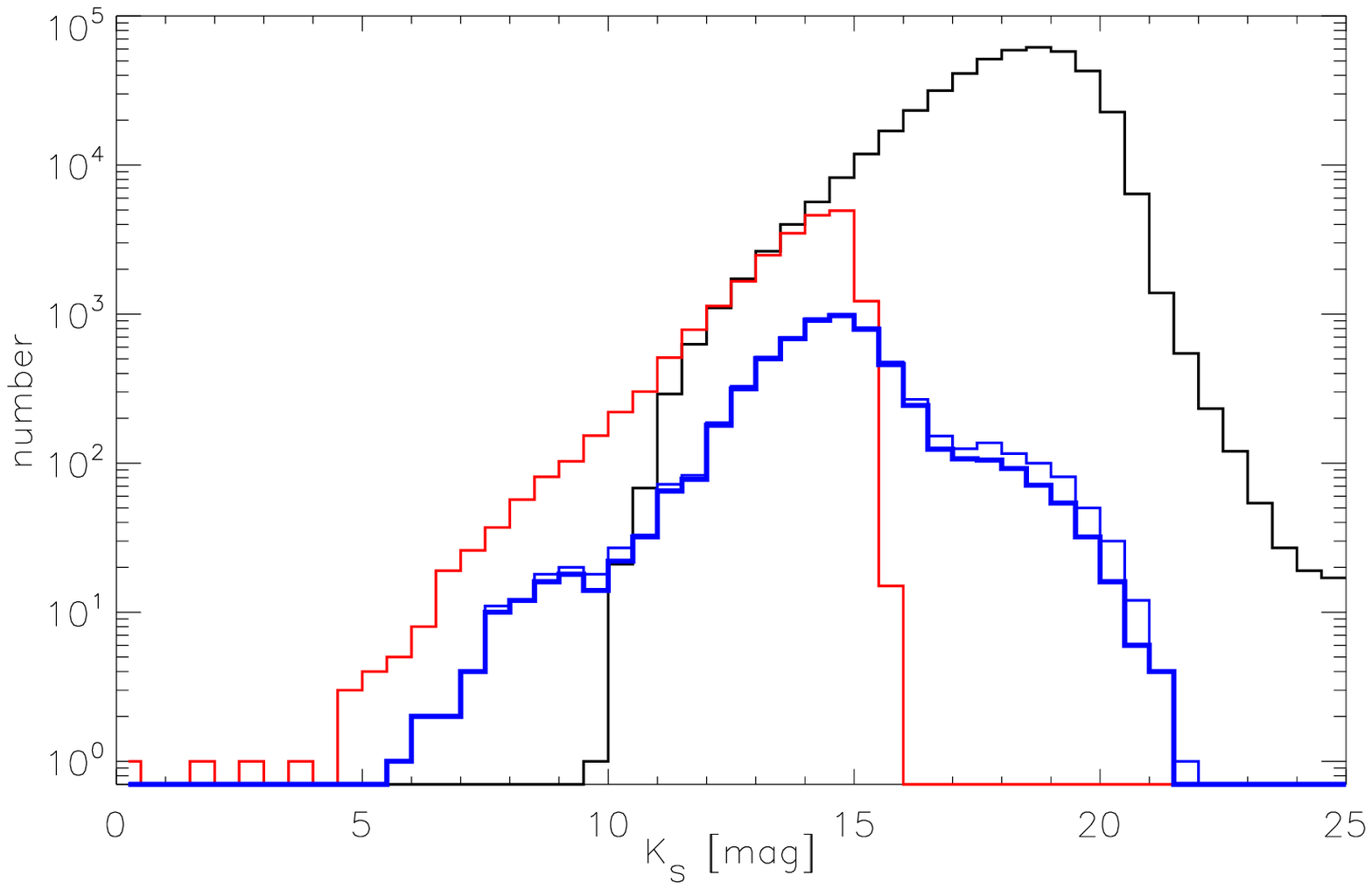}
\figcaption{Histograms of the infrared magnitudes of objects
inside the HAWK-I field.
The red line shows all 2MASS sources, the black line all HAWK-I sources,
the thin blue line all X-ray sources with an infrared counterpart, and the
thick blue line the X-ray sources classified as Carina members (class = H2).
\label{histo.fig}}
\end{figure}

Objects as faint as
$J \sim 23$, $H \sim 22$, and $K_s \sim 21$ are detected with S/N~$\ge 3$.
However, since significant parts of the CNC are pervaded by
very strong and highly variable diffuse emission that reduces the
local source detection sensitivity,
it is not possible to
determine a strict and unique value for a completeness limit valid
for the full area of the {HAWK-I} survey.
An estimate for the  ``typical'' completeness limit
(i.e.~the limit above which we can expect most objects in the
survey area to be detected) can be derived from the histograms of the
magnitudes\footnote{Although the turnover in the source count curves is not 
a formal measure of completeness, it can serve as a proxy to show the 
typical values of the completeness limit across the field. 
}
 (see Fig.~\ref{histo.fig}) and yields
$J_{\rm compl} \sim 21$, $H_{\rm compl} \sim 20$, and $K_{s,\, \rm compl} \sim 19$.
Nearly all objects above these limits are S/N~$\ge 10$ detections.

The total number of detected objects is highest in the $H$-band images
and smaller in the $J$-band (mainly due to extinction) and
in the $K_s$-band (mainly due to the strong diffuse nebulosity).
Finally, a merged HAWK-I photometric catalog was generated from 
the source lists of all 24 mosaic positions and all three photometric bands;
duplicates from the overlap regions between the individual mosaic
positions were removed by keeping the catalog entry with the highest signal-to-noise ratio.

\subsection{The HAWK-I source catalog}

The final HAWK-I photometric catalog contains 600\,336 individual objects.
This is about 20 times more than the
number of 2MASS sources in the area of the HAWK-I mosaic (29\,218).
Most (502\,714) catalog objects are simultaneously detected
 in the $J$-, $H$-, and the $K_s$-band.
Since the main aim of the HAWK-I study was to detect the faintest
possible objects,
the catalog contains numerous very weak objects close to the detection 
limit,
some of which may be spurious false-detections.
However, the objects that are detected independently \textit{in all three
NIR bands} are very likely real detections.
The significance of the infrared source detection can be quantified by the
formal photometric uncertainties\footnote{What we denote as
``formal photometric uncertainties'' are errors reported
by the photometry software. In addition to these
random errors, the total photometric uncertainty also includes
systematic error components in the HAWK-I data and calibration
errors, as discussed below.}
 in the following way:
the magnitude limits below which more than 90\% of the listed sources
have formal photometric uncertainties
less than 0.1~mag (corresponding approximately 
to $S/N \ge 10$ detections)
are $J \le 21.2$, $H \le 20.3$,
and $K_{\rm s} \le 19.3$. 
On the other hand, 
the magnitude limits above which more than 90\% of the listed sources
have formal photometric uncertainties
larger than 0.3~mag (corresponding approximately 
to $S/N \le 3$ detections)
are $J \gtrsim 23$, $H \gtrsim 22$,
and $K_{\rm s} \gtrsim 21$.

Finally, we checked
the integrity of the derived photometric data in the 
HAWK-I catalog
by comparing the HAWK-I and 2MASS photometry for a 
carefully selected sample of
stars that are (1) detected with S/N $> 10$ in 2MASS,
(2) faint enough to avoid saturation problems in
the HAWK-I data, (3) isolated enough
to have no other infrared source in the HAWK-I images within
$3.5''$,
and (4) have formal photometric uncertainties 
of $\leq 0.1$~mag in each band.
For the 714 stars in this sample we determined
the standard deviations between HAWK-I and 2MASS
magnitudes and colors. Applying ``$3\sigma$-clipping''
to remove obvious outliers, we found
$\sigma_J = 0.11$~mag, $\sigma_H = 0.08$~mag, $\sigma_{K_s} = 0.11$~mag,
$\sigma_{J-H} = 0.12$~mag, and $\sigma_{H-K_s} = 0.11$~mag.
This level of accuracy appears quite reasonable if we consider
the numerous sources of uncertainties.
Firstly, the accuracy of the photometric calibration is
naturally limited by the photometric
quality of 2MASS. 
The mean photometric errors listed for the 2MASS stars
suitable for calibration are
$\langle \sigma_J \rangle_{\rm 2MASS} = 0.05$~mag, 
$\langle \sigma_H \rangle_{\rm 2MASS} = 0.04$~mag,
and $\langle \sigma_{K_s} \rangle_{\rm 2MASS} = 0.05$~mag. 
Therefore, a large part
of the total photometric uncertainties originates from
the 2MASS data.
Secondly, we are dealing here with a wide-field mosaic of
a region pervaded by strong and highly variable diffuse nebulosity,
that was observed during a number of nights with slightly variable
conditions.
Thirdly, we note that the accuracy to which
such a dataset can be photometrically calibrated
is limited by a number of effects, including
variations in the pixel spatial scale across the camera,
geometrical/optical effects such as vignetting,
non-uniformity in the flat illumination,
chip-to-chip quantum-efficiency color effects,
non-linearities in the detectors, and color differences in the
illumination of the flats and the average astronomical objects.
Although each of these effects is quite small, their combination will
undoubtedly cause systematic spatial photometry effects
at the level of several percent,
particularly around the edges of the detectors. 
In our data set, such spatial effects seem to produce 
systematic uncertainties at levels of 
up to $\sigma \lesssim 0.15$~mag.
Although this would be problematic if one would try to 
use color-selection criteria to define samples of
young stars, it is a much less serious problem for our analysis
of the X-ray selected sample of objects.
For the present study, this level of photometric
accuracy is sufficient for our scientific purposes.

\subsection{Background contamination in the HAWK-I catalog\label{imf_expectation}}

Although we can expect that our HAWK-I data
detect essentially all young stars in the survey area,
these infrared data alone do not allow reliable
distinction of young stars
from the numerous background objects.
Simple extrapolations, based on the known population of high-mass stars
in the CNC, show that most of the
objects in the HAWK-I catalog must be
background sources, unrelated to the CNC.
\citet{SB07} estimate that the total stellar population
of the CNC consists of about 65\,000 stars.
About 40\,000 of these should be located in the 
HAWK-I field-of-view.
This suggests that {\em the vast majority  ($\gtrsim 93\%$) of the 
$\sim 600\,000$ detected infrared sources in our HAWK-I catalog
must be foreground or background sources}.
This estimate agrees well with the results 
from \citet{Povich11a}, who found
that 93.4\% of all infrared sources seen by  \textit{Spitzer}
in the CNC are probably unrelated field stars.
Considering the above described problems with infrared-excess selected
samples of young stars, we conclude that
the HAWK-I data alone can yield only very limited
insight into the young stellar populations in the CNC;
only a combination of X-ray and infrared data can
yield a reasonably clean and complete sample of young stars.

\section{Infrared counterparts of the X-ray sources}

\subsection{Catalog matching}

We employed the method described by 
\citet{Broos11a}, that uses positional uncertainties for individual 
sources plus a match significance threshold,
to match the X-ray and infrared catalogs 
in the CNC. 
In the first stage, the algorithm tests the hypothesis that a possible 
pair of sources from the two catalogs is spatially coincident.
The most significant match of each master source is referred to as its 
``Primary Match''; any other
significant matches are ``Secondary Matches''.
The second stage of the algorithm resolves possible 
many-to-one and one-to-many relationships between the X-ray 
``master catalog'' and the infrared ``slave catalog''.
Clear one-to-one relationships are classified as ``successful primary matches'',
while in cases where, e.g., two master sources are significantly close to
a single slave source, the less significant primary match
 is labeled as ``failed''.
This finally provides a reasonable one-to-one set of matches.
We point out that no source-matching method can produce
100\% perfect results;
both false negative
(true physical associations that are missed) and false positive
(unphysical associations that are listed) are likely present,
but their numbers should be quite small.

The area of the HAWK-I mosaic
covers 27\% of the area of the {\it Chandra} survey
and contains 52\% (7472) of all 14\,368 {\it Chandra} sources. Only
3620 (48.4\%) of these 7472 {\it Chandra} sources in the HAWK-I field
have successful primary matches with 2MASS counterparts, and just
2534 (33.9\%) have valid $J$- {\em and} $H$- {\em and} $K_s$-band photometry 
from 2MASS.
The matching of the {\it Chandra} X-ray catalog to the 
HAWK-I infrared catalog improves these numbers substantially:
For 6583 of the 7472 Chandra sources in the HAWK-I field
successful primary matches are found. 
Among these are 3016  {\it Chandra} sources that had no 2MASS matches.
The matching with our HAWK-I catalog 
increases the number of known infrared counterparts to the
{\it Chandra} sources in the HAWK-I field to  6636  (88.8\%),
nearly two times more than based on the 2MASS catalog alone.

In the probabilistic source classification described in \citet{Broos11b},
6173 (93.0\%) of these 6636 X-ray sources with infrared counterparts
are classified likely Carina members,
131 (2.0\%) as likely foreground stars,
none (0\%) as likely background stars,
66 (1.0\%) as likely AGN, and 266 (4.0\%)  remain unclassified.

For 836 (11.1\%) of all X-ray sources in the HAWK-I field
no matching infrared counterpart is found in the HAWK-I catalog. 
31 (3.7\%) of these are classified as likely foreground stars,
626 (74.7\%) as likely Carina members, none (0\%) as likely
background stars,
41 (4.9\%) as likely AGN, and 138 (16.6\%) remain unclassified.
In about half of these cases,
a visual inspection of the HAWK-I images
shows no perceptible infrared source at 
the X-ray source position. Besides
extragalactic sources that are extremely
faint in the near-infrared,
some X-ray detected young stars in the CNC
may remain undetected in the HAWK-I data
due to very strong extinction\footnote{
A $0.1\,M_\odot$ star with an age of 3~Myr should have $K_s = 17.3$;
for $A_V \ge 15$~mag such a star would be fainter than the typical
completeness limit of the HAWK-I data and might be undetected in the
near-infrared. In the X-ray band, however,
young stellar objects with optical extinctions of several hundred magnitudes
have been detected, 
\citep[see, e.g.][]{Grosso05}.}
 or because of very strong local nebulosity.
Finally, in some cases, the X-ray source may be spurious;
we note that 242 (28.9\%) of the 836 X-ray sources in HAWK-I field without 
infrared match have $\leq 3$ net counts  in the {\it Chandra}
data.
In the other half of these cases, an infrared source is seen 
close (within $\sim 2''-3''$) to the X-ray source position,
but just outside the matching region. 
Some of them may be ''false negative'' rejections
of physical matches. 
We decided {\em not} to add matches ``by hand'' in such cases,
in order to keep the matching as objective as possible.

\subsection{Secondary matches \label{sec-matches.sec}}

A substantial number (1184 of 6583, i.e.~18\%) of the {\it Chandra} sources in the HAWK-I 
field with infrared matches have more than one possible counterparts
within the search radius. Any of these matches qualifies
as a counterpart to the X-ray source, and thus we
 have to decide which of these matches
(the primary match or one of the secondary matches)
we consider to be the most likely true counterpart.
In the automatic source matching, this decision is purely based
on the angular separation between the positions of the
X-ray sources and the possible matches. This is a reasonable
choice in the absence of further information, but 
may, of course, not always be correct.
Due to the very high density of very faint objects in the HAWK-I catalog, 
some of the matches with very faint infrared sources may in fact 
be chance superpositions of physically unrelated sources, and one of the 
``secondary matches'' may be the true counterpart.

We note that some a-priori information about the
likely nature of the X-ray sources is actually available and can be used 
as a guide for this decision. Since we are looking
at a star forming region, we can clearly expect that
many of the X-ray sources are young stars in Carina, most of which should
be relatively bright infrared sources.
The statistical analysis of possible X-ray contaminants
\citep{Getman11} actually shows that the number of expected 
infrared-faint background (stellar and extragalactic) contaminants is 
much smaller than the number of detected X-ray sources.
Thus, a given X-ray source in the Carina Nebula is a-priori more likely 
to be a young star than a background object.
This is relevant for cases where 
the primary match of an X-ray source is a very faint infrared source
 (especially if it is below the expected infrared brightness for X-ray detected
Carina members, 
i.e.~$H \gtrsim 19$) {\em and} a significantly brighter secondary match 
is present. In these cases,
the brighter secondary match can be
more likely to be the true physical counterpart to the X-ray source\footnote{
The following ``Gedankenexperiment'' may illustrate the situation:
imagine that we are searching for the counterpart of a specific 
X-ray source in a sequence of deeper and deeper infrared images, and
assume that the nth image in the sequence reveals 
the true physical counterpart of the X-ray source.
Due to the unavoidable uncertainties of the position determinations,
the angular offset between the X-ray and infrared positions is
not exactly zero but some (small) fractions of the search radius.
The subsequent, deeper infrared images will detect an ever increasing
number of very faint infrared sources. It will thus get more and
more likely that one of these very faint infrared sources
falls randomly very close to the X-ray source position 
and edges out the true physical match with the brighter infrared source.
The frequency of unphysical random matches depends on the
number density of the faint infrared sources, i.e. increases
strongly with the depth of the data set used for X-ray source matching.
It is very low for typical medium-deep data (such as
 2MASS), but increases strongly for very deep
infrared data such as our HAWK-I survey. 
}.

A statistical estimate of how often one of the very numerous very faint
infrared sources should fall randomly into the match region
of an X-ray source
and thus {\em might displace an otherwise perfectly valid match with a
brighter infrared source}, can be made as follows:
The HAWK-I catalog contains 250\,790 very faint sources with $H>19$ 
in the 1280 square-arcmin field, i.e.~the
mean number of these faint sources is 0.051 per square-arcsecond.
Since  the median X-ray error circle radius
is $0.32''$ \citep[see][]{Broos11a},
Poisson statistics shows that the probability
to have one or more purely random match with such very faint 
infrared sources is 0.017. 
We thus expect that about 130 of the 7472 X-ray sources in the HAWK-I field
should have one or more unphysical random matches with very faint
infrared sources.

This estimate is in good agreement with the results of the source
matching:
For 115 {\it Chandra} sources one of the secondary matches
is brighter (in the $H$-band) than the primary match.
Since there is a high probability that these
very faint primary matches are unphysical random associations, 
we performed a detailed visual inspection of these sources.
We found 74 cases where one of the secondary matches is considerably brighter
than the very faint primary match, and its position is very well
consistent with that of the X-ray source; in these cases,
we replaced the original
very faint primary match by this brighter secondary match.
The remaining cases were unclear: often, the
X-ray sources are
located just between two similarly bright infrared sources,
only marginally closer to the slightly fainter source.
In these unclear cases, we kept the identification with the
primary match.
 
To conclude the justification for this procedure, we note
that if we would have used less deep infrared images 
(such as the 2MASS data) for
the identification of the X-ray sources, the very faint
primary matches (some of which we have deleted) would not have been
detected and thus the brighter secondary match
(which we prefer in some cases) would have been regarded as the
primary match from the beginning.

\subsection{Final selection of NIR magnitudes for the {\it Chandra} sources}

For our final infrared counterpart catalog we generally preferred
the HAWK-I magnitudes.
The 2MASS magnitudes were used for 360 X-ray sources,
including  bright objects that are above or close to
the expected saturation limit of the HAWK-I images, 
and cases where an X-ray source has a valid 2MASS match 
and its HAWK-I match is very faint\footnote{The justification for this
is analogous to the arguments given in Sect.~\ref{sec-matches.sec}}
($H \ge 18$).

The central region of the dense cluster Tr~14 requires
special attention. The stellar density there is so high that even
our HAWK-I images are affected by crowding and source blending,
which may lead to unreliable photometry for some objects.
\citet{Ascenso07} imaged a $\sim 1' \times 1'$ area centered
on the cluster core with the
near-infrared adaptive optics instrument NACO.
With an angular resolution of $\sim 0.1''$ these
AO images resolved the cluster core and allowed them to derive
$J$-, $H$-, $K_s$-photometry for about 180 stars
in their field-of-view.
Matching the {\it Chandra} catalog to the NACO source table 
\citep{Broos11a} yields
54 successful primary matches,  45 of which have NACO photometry
in all three bands.
We therefore used the NACO photometry for these 45 {\it Chandra}
source matches in the center of Tr~14 instead of the 2MASS or HAWK-I
photometry.
This yields $J$-, $H$-, $K_s$-photometry for 8 X-ray sources
that had neither 2MASS nor HAWK-I counterparts.

In our final combined catalog, which is given in Table~\ref{magnitudes.tab},
 6241 X-ray sources in HAWK-I field
have valid
photometry in all three  $J$-, $H$-, and $K_s$-bands.

%
\section{Results\label{results.sec}}

\subsection{Infrared magnitudes of the X-ray sources}

In Fig.~\ref{histo.fig} we compare histograms of the infrared magnitudes
of all 2MASS and HAWK-I sources in the HAWK-I field to
those of the X-ray selected sources in this area.
The peak of X-ray selected sources is at $J \sim 16$,
$H \sim 15$, and $K_s \sim 14.5$. This agrees well with the
expected NIR magnitudes of the faintest detectable X-ray sources
discussed above.
A few hundred X-ray sources have very faint 
infrared counterparts with magnitudes $\gtrsim 20$;
many of these are likely (extragalactic) background contaminants,
but some may be flaring very low-mass stars or deeply embedded 
young stellar objects.

\subsection{Near-infrared excesses of the X-ray sources}

\begin{figure}[h] \epsscale{0.5}
\plotone{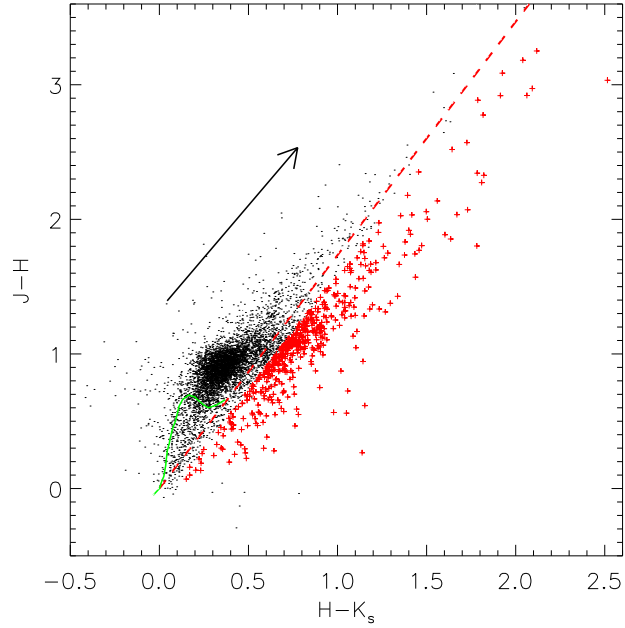}
\figcaption{Color-color diagram for the X-ray selected objects
in the HAWK-I field.
The solid line shows the main sequence,
the arrow shows the $A_V = 10$ mag reddening vector with slope
1.73,
and the dashed line marks the red edge of the
photospheric reddening band.
Objects are classified as near-infrared excess sources if
they lie at least 0.05~mag to the
right and below the reddening band and above $J-H=0$.
Near-Infrared excess sources  are shown as red crosses.
\label{ccd.fig}}
\end{figure}

In Fig.~\ref{ccd.fig} we show the $J-H$ vs.~$H-K_s$ 
color-color diagram for the X-ray selected sources.
We find that 485 of the 
 6241 X-ray sources in HAWK-I field with valid
photometry in the $J$-, $H$-, and  $K_s$-band show
near-infrared excess. The global excess fraction of the
X-ray selected sample of infrared sources  in HAWK-I field 
is thus $(7.8\pm 0.3)\%$.

In order to compare the excess fraction to other star forming
regions, we have to take into account that the CNC contains
clusters with different ages \citep[we adopt the age estimates
as given in][]{SB08}. We use the results of the clustering analysis of
\citet{Feigelson11} 
to define the membership of the individual X-ray sources
to the different clusters or the distributed population.
Note that further and more comprehensive studies
of some of these sub-populations of the CNC are given
in other publications in the frame of the CCCP
\cite[see][]{Wang11,Wolk11}.
 
\textit{Tr~16:}
According to the results of the clustering study,
Tr~16 consists of seven sub-clusters that 
contain a total of  530 X-ray sources that are classified as
likely Carina members.
For 449 of these we have $J$- and $H$- and $K_s$-band magnitudes, and
 31 of these sources show near-infrared excess, resulting 
in an near-infrared excess fraction of $(6.9\pm 1.2)\%$.
 This value is clearly lower than the
near-infrared excess fraction expected for a $\sim 3$~Myr old population 
\citep{Briceno07}.

\textit{Tr~14:}
The clustering analysis related 1378 {\it Chandra} sources to Tr~14,
for 1219 of which we have complete NIR photometry.
118 of these objects have near-infrared excesses; with $(9.7\pm 0.8)\%$,
the near-infrared excess fraction is slightly higher than in Tr~16, but 
again lower than typical for $\sim 1-2$~Myr old populations. 

\textit{Tr~15:}
The clustering analysis related 481 {\it Chandra} sources to Tr~15, 
for 436 of which we have NIR photometry; only
9 of these have near-infrared excesses. The resulting
near-infrared excess fraction of $(2.1\pm 0.7)\%$) in 
Tr~15 is thus clearly lower than for Tr~16 and Tr~14.

\textit{Treasure Chest:}
There are 96 {\it Chandra} sources associated with the very young
($\lesssim 1$~Myr)``Treasure Chest'' cluster.
25 of the 78 X-ray detected cluster members for which we have NIR photometry
display near-infrared excesses. The corresponding near-infrared excess fraction
of $(32.1\pm 5.3)\%$ is clearly much higher than for the other
clusters in the CNC.

\textit{The ``widely distributed'' population:}
According to the clustering analysis of \citet{Feigelson11},
5185 of the X-ray sources are not associated with any cluster
and thus belong to a widely dispersed
population of young stars, extending throughout the  CNC.
For 1412 of these objects in the field of the HAWK-I survey
we have complete $J$- and $H$- and $K_s$-band photometry, and
94 of these stars have near-infrared  excesses.
The near-infrared excess fraction of $(6.7\pm0.7)\%$ is slightly lower
but still similar to that in the
clusters Tr~16 and Tr~14.

\subsection{Extinction of the X-ray sources\label{extinctions}}

In principle, the
location of a specific source in the color-color diagram 
can be used to infer its extinction by shifting it ``back''
to the intrinsic (un-reddened) color along the direction of the
reddening vector. Since the intrinsic colors of the sources
depend on their spectral type, which we do not know for most objects,
we cannot determine accurate extinction values of individual objects.
Nevertheless, we can derive
statistical estimates of the distribution of extinction for
the whole sample of objects in the
following way: We assume a ``typical'' intrinsic color of $H-K_s = 0.1$, 
which is roughly correct for K-type stars\footnote{Note that, at an age
of $\sim 1-3$~Myr, a solar-mass pre-main sequence star has a spectral 
type of $\sim K7$.}
 which are thought to dominate the X-ray selected sample. 
In order to compute an estimate of the visual extinction, we 
take the apparently anomalous extinction law in Carina into account:
as discussed in more detail in \citet{Povich11b},
the average extinction law toward the CNC stars is better represented by 
$R_V = 4$ instead of the standard diffuse interstellar medium value of 
$R_V = 3.1$. According to the relations given in \cite{Mathis90}, 
the formula connecting color excess to visual extinction is then
$\tilde{A_V} = (H-K_s - 0.1) \times 13.7$.
We restrict this analysis to sources within the photospheric reddening 
band, since for sources with near-infrared excesses
the colors are no longer proportional to the extinction 
\citep[e.g.,][]{Stark06}.
Furthermore, we exclude all objects with photometric uncertainties larger
than 0.1~mag in any of the near-infrared bands and all objects 
classified as background stars and AGN.
By using the notation $\tilde{A_V}$  we want to emphasize
that these values do {\em not} accurately reflect the true extinction 
for {\em individual} objects; however, the distribution 
of $\tilde{A_V}$ values should yield a reasonably good proxy for the
distribution of the true extinctions, since 
most errors are expected to cancel out in this large sample.

\begin{figure}[h] \epsscale{0.5}
\plotone{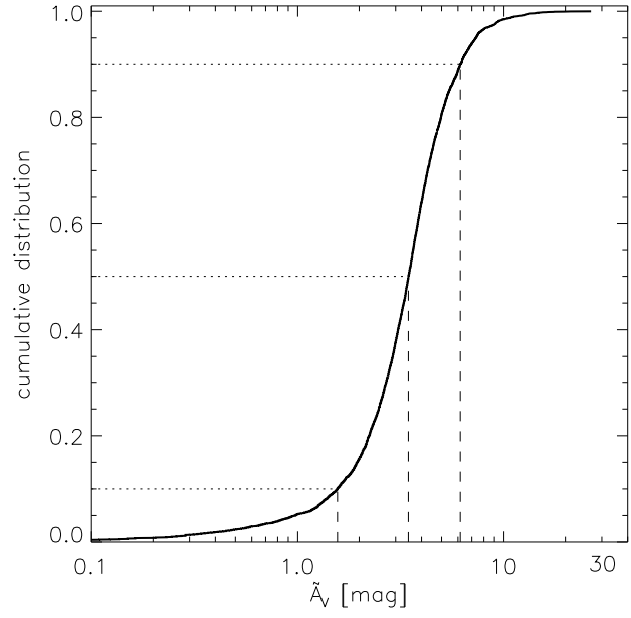}
\figcaption{Distribution of extinction estimates for the X-ray selected
sources in the HAWK-I field of the CNC. The dashed/dotted lines mark the
10\%, 50\%, and 90\% quantiles.
\label{av-distr.fig}}
\end{figure}

The resulting distribution of extinction estimates for 4879 objects
(see Fig.~\ref{av-distr.fig}) has a median value of 
$\tilde{A_V} \sim 3.5$~mag and extends up to $\tilde{A_V} \sim 27$~mag.
The 10th and 90th percentile values are 1.6~mag and 6.2~mag, respectively.
This variation by a factor of nearly 4 implies that there is
considerable differential extinction between the individual young stars
in the region.
Note that this range is in good agreement with the above mentioned
estimates of the cloud extinction ($A_V \sim 1 - 10$~mag); this suggests that the
X-ray selected stars fill the whole range of cloud depths, i.e.~we
do not systematically miss a population of
highly obscured members.

Finally, we emphasize again that these extinction estimates are
only valid for the sub-sample of objects {\em without near-infrared 
excess} (i.e.~diskless stars). Among the objects with
near-infrared excesses,  stars with very red color suggesting
higher extinctions ($A_V \ge 10$~mag) are clearly present.
However, the relative number of X-ray detected objects with very red colors is
considerably smaller than seen in many other star forming regions:
only 29 (0.5\%) out of 6241 X-ray selected objects have $H-K_s \ge 1.5$.

We conclude that the vast majority of the X-ray detected young stellar
population in the CNC shows rather moderate extinction, 
$A_V \le 10$~mag;
the X-ray selected Carina members can thus be considered as a ``lightly obscured''
population of young stars.
In this context, it is important to note that the study of
\cite{Povich11a}, which is based on deep \textit{Spitzer} mid-infrared
images of the CNC, found a considerable population (1439) of young star candidates
with strong mid-infrared excesses, most of which are {\em not} detected
in the {\it Chandra} X-ray data. This non-detection seems to be related to
the relatively strong obscuration of many of these objects.

\subsection{Color-Magnitude Diagrams of the X-ray selected objects \label{contamination.sec}}

\begin{figure}[h] \epsscale{1.0}
\plottwo{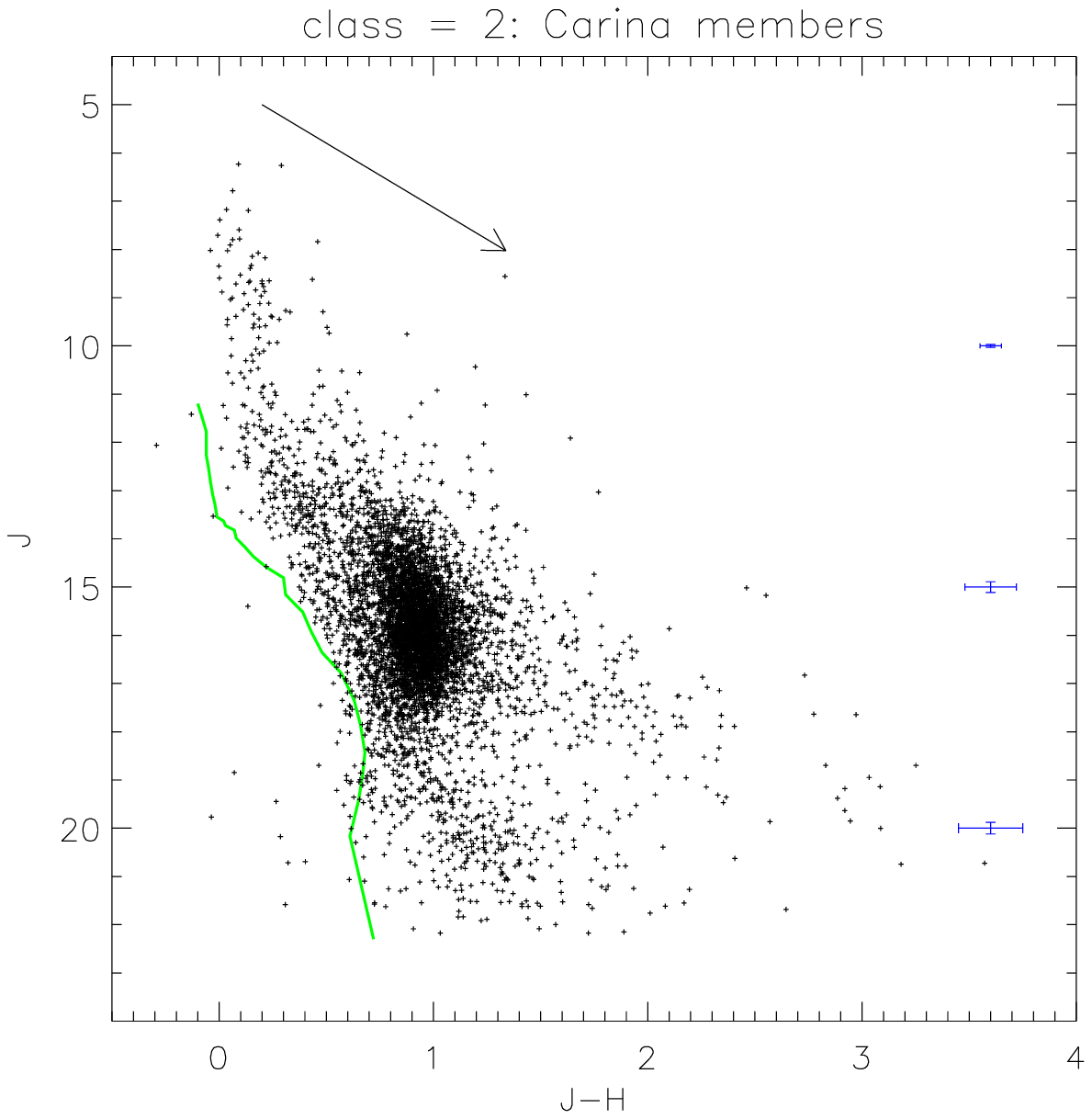}{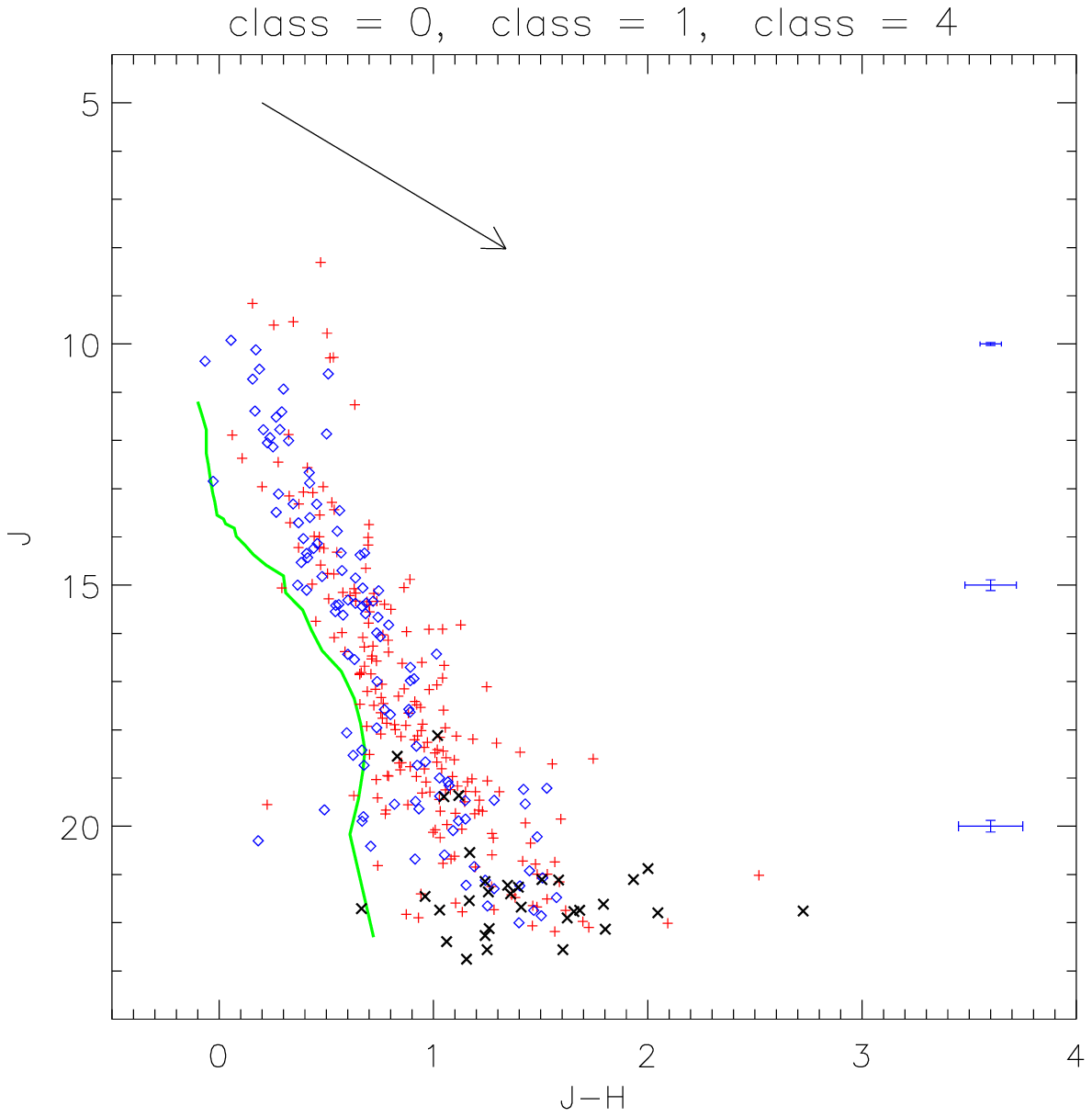}
\figcaption{Left: Color-Magnitude Diagram of X-ray selected objects
classified as likely Carina members (class = H2).
Right: Color-Magnitude Diagram of X-ray selected objects
classified as likely non-members (foreground stars are marked
by blue diamonds, AGN by black {\sf X} signs, and unclassified objects
as red crosses).
In both plots, the
solid line shows the ZAMS from \cite{Siess00} for a distance of 2.3~kpc;
the arrows indicate reddening vectors for $A_V = 10$~mag.
Typical uncertainties of the photometry for different magnitude ranges
are indicated by the
sequence of three errorbars near the right edge.
\label{cmds-members.fig}}
\end{figure}

In Fig.~\ref{cmds-members.fig}a we show the $J$ versus $J-H$  
CMD of the X-ray selected objects that
were classified as likely Carina members (class = H2).
The stars at the top of the diagram, i.e.~at $J \leq 10$, are O- and early
B-type stars which are known to be strong X-ray emitters. Going towards
fainter magnitudes, the CMD shows a ``gap'' (i.e.~a lower number of objects) 
in the range $J\sim 10-11$; this corresponds to stars with late B and 
A spectral types, most of which are not expected to be strong X-ray sources.
Proceeding to fainter magnitudes, the number of objects increases strongly,
as expected for a mass function rising towards lower masses.
The vast majority of the objects 
are at CMD locations corresponding to stellar masses of
$\gtrsim 0.25\,M_\odot$, as expected from the X-ray detection limits. 
The objects with very faint magnitudes
($J \gtrsim 19$) may be very-low mass stars with 
unusually high X-ray luminosities (perhaps due to X-ray flares), 
or objects that are surrounded
by circumstellar disks seen approximately edge-on.

As the probabilistic source classification cannot provide definitive
evidence for the nature of individual objects, we
use this diagram to search for indications of remaining contamination
among the objects classified as likely Carina members.
The first thing to note is that nearly all objects lie to the right of
the zero-age main sequence (ZAMS), i.e.~at a location where one would expect 
young pre-main sequence stars in Carina.
However, 46 of the 6230 objects in the diagram lie to the
left of the ZAMS, and another 90 objects on or within  0.1~mag
of the ZAMS. These objects may actually be mis-classified fore- or background 
objects, suggesting a low level ($\sim 2\%$) of field star contamination
in the X-ray selected sample.
Further indications of contamination are seen at the faintest magnitudes.
If X-ray detected young stars are very faint ($J \geq 19$) because they
suffer from strong inter- or circumstellar extinction, they should
show correspondingly red colors ($J-H \ge 2$).
However, the CMD contains some $\sim 270$ very faint objects 
that have only moderately red colors ($J-H < 2$); many of these objects 
may actually be mis-classified AGN.

Looking at the CMD for the objects that were classified as
likely non-members (Fig.~\ref{cmds-members.fig}b), 
we see that likely
galactic foreground objects are at CMD locations 
consistent with slightly reddened main-sequence stars,
whereas the likely extragalactic
(AGN) objects are strongly concentrated towards very faint
magnitudes. Again, we conclude that the CMD positions of the
objects are consistent with their automatic classifications.
Although it is possible that some of the objects classified
as background are actually Carina members \citep[see][]{Broos11a}, 
their number is too
small to  significantly affect the source
statistics. There is not much we can conclude about the
unclassified sources; some
fraction of these may actually be Carina members, but
again their number is rather small.

In summary, we find that the automatic source classification
produced very reasonable results and the remaining level of contamination 
among the objects classified as Carina members (class = H2) is very low.
Even if we assume that most (i.e.~$\sim 250$) of the very faint 
class = H2 objects
are actually mis-classified AGN, the total number of mis-classifications
(i.e.~250 plus the $46 + 90$ objects to the left or close to the ZAMS)
would be only $\approx 400$, i.e.~just 6.4\% of all class = H2 objects
in the HAWK-I field.
From this we conclude that
a maximum of $\approx 7\%$ of X-ray sources with H2 classifications may in fact 
not be Carina members.
We therefore will from now on focus on the infrared properties
of the objects classified as Carina members and assume that
these constitute a reasonably complete (but of course
X-ray luminosity-limited, i.e.~approximately mass-limited) and 
only very marginally contaminated sample of the young stellar 
population in the CNC.

\subsection{The $K$-band luminosity-function of X-ray sources}

\begin{figure}[h] \epsscale{0.5}
\plotone{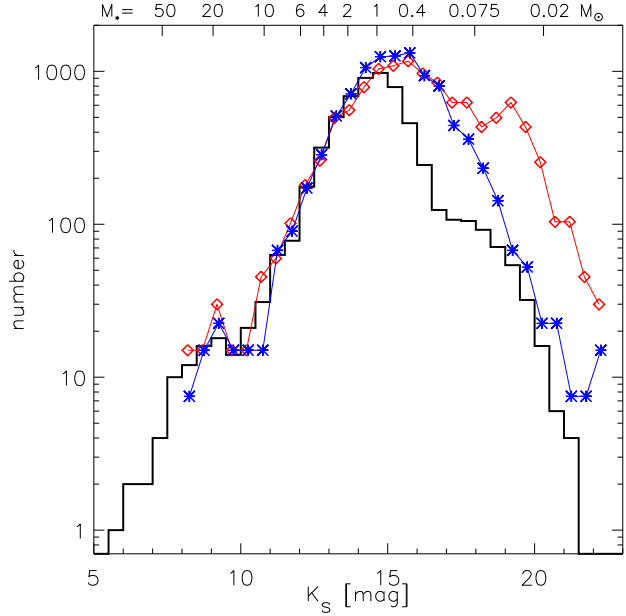}
\figcaption{The $K_s$-band luminosity function (KLF) of all \textit{Chandra} sources
in the HAWK-I field that are classified as likely Carina members
(black histogram) is compared to the KLF of the Orion Nebula Cluster
derived from the infrared observations by \citet{Muench02} (red line
with diamonds, shifted by 3.69~mag to account for the difference in
distances and multiplied by a factor of 15 for normalization),
and the KLF for the X-ray detected Orion Nebula Cluster members
from the \textit{Chandra} Orion Ultradeep Project \citep{Getman05}
(blue line with asterisks). On the upper x-axis we have marked
approximate stellar masses corresponding to the magnitudes
on the lower x-axis, based on the (pre-main sequence-) stellar models
form \citet{Baraffe98} for the mass range
0.02--0.5~$M_\odot$, \citet{Siess00} for the
mass range 0.5--7~$M_\odot$, and \citet{Lejeune01} (model
 iso-c020-0650) for the mass range 7--70~$M_\odot$,
and assuming an age of 3~Myr and an extinction
of $A_V = 3.5$~mag.
\label{klf.fig}}
\end{figure}

Information about the distribution of stellar masses of the 
X-ray selected objects in the CNC can be inferred from the
$K_s$-band luminosity function (KLF).
In Fig.~\ref{klf.fig} we compare the KLF
of the X-ray selected objects classified as
Carina members in the HAWK-I field to the
KLFs of the infrared sources and the X-ray selected young stars
in the Orion Nebula Cluster. Up to magnitudes of $K_s \lesssim 15$,
i.e.~the level above which we expect the X-ray selected Carina sample to be nearly
complete,  the Carina and Orion KLFs agree very well.
This implies that for masses of $M_\star \gtrsim 1\,M_\odot$
the shape of the IMF in the CNC is consistent with the IMF in the 
Orion Nebula Cluster (which, in turn, is known to agree with the field IMF).
At fainter magnitudes, for $K_s \gtrsim 15$, the steep drop in the
number of X-ray detected Carina members reflects the
expected incompleteness of the  X-ray selected Carina sample arising
from the X-ray completeness limit (which is roughly at 
$\log L_{\rm X, lim} \sim 30.2$~erg/s for Carina,
 compared to $\log L_{\rm X, lim} \approx 27.5$~erg/s for the ``lightly absorbed
optical sample'' in the Orion data).

For even fainter magnitudes, $K_s > 17$, the Carina KLF does not
drop as steeply as the Orion KLF of X-ray selected stars,
but seems to show a ``bump'' in the histogram, i.e.~a surplus of very
faint objects. The stellar masses corresponding to these magnitudes
would be $\lesssim 0.2\,M_\odot$, i.e.~below the
estimated X-ray detection limit. Although some of these faint objects
might be optically strongly obscured, highly X-ray luminous 
young stars, at least some fraction of them
might be mis-classified non-members of the CNC.
The size of this ``faint surplus population'' is, however, rather small;
by extrapolation of the histogram slope in the $K_s = 15 - 17$~mag interval
we find that it comprises just $\sim 300$ objects. 
This number agrees well with the estimate made above that some
$\sim 200-300$ of the faintest objects may in fact be mis-classified
background objects.

In summary, our analysis of the KLF does
not reveal any indications for systematic differences between
the shape of the IMF in the CNC 
and the Orion Nebula Cluster (in the mass range $\sim 1 \dots 20\,M_\odot)$.
This suggests that
the shape of the IMF in the CNC is (at least down to $\sim 1\,M_\odot$)
consistent with the field IMF.
This conclusion relies on the assumption that the X-ray
luminosity function (XLF) of the CNC stars is similar to that
of the Orion stars.
This assumptions appears reasonable, because  X-ray studies of a large number
of young stellar clusters have shown that the XLF appears to be universal 
\citep[see][]{FeigelsonIMF}. Furthermore,
the analysis of the CCCP data in \citet{Feigelson11} has directly shown that 
the Carina XLF has (above the completeness limit) a 
very similar shape to the Orion XLF.


\section{X-ray clusters and infrared clusters}

\citet{Feigelson11} investigated the
clustering properties of the X-ray sources classified as
likely Carina members and
identified 20 principal clusters of X-ray stars as well as 
31 small groups of X-ray stars outside the major clusters.
While most of the principal clusters are clearly evident
in optical images, the reality of most of the apparent
small groups of X-ray stars is not completely sure; some of them
could perhaps just be random fluctuations in the
spatial X-ray source density.
To investigate the reality of these clusterings, we inspected
our HAWK-I images at the positions of all X-ray clusters and groups.
The advantage of the HAWK-I images is that they show 
essentially all young stars (except some very deeply embedded objects),
whereas the flux limit of the {\it Chandra} data implies that only 
$\sim 30\%$ of the stellar cluster members are
detected as X-ray sources.
However, the HAWK-I images have the disadvantage of
a much higher level of background contamination, that may 
easily prevent a loose clustering of sources from being
recognized against the dense and irregular background of 
contaminating infrared sources.

In our inspection we found that
all but one of the 20 principal X-ray clusters are related
to obvious clusterings in the HAWK-I images.
The exception is cluster 19 (104559.5-600832), for which
no indication of a clustering can be seen in the HAWK-I images;
we note that the X-ray sources in this cluster show unusually high
median photon energies, suggesting that the objects are particularly
strongly obscured, which may explain the non-detection in the HAWK-I
images. This is supported by the fact that this
X-ray cluster coincides with a clustering
seen in the \textit{Spitzer} images
\citep[see][]{Smith10b}.
While the near-infrared confirmation of the X-ray clusters is
trivial for the well known
optical clusters,
it is relevant for the small clusters 13, 14, 16, 17, 18, and 20.
For the 31 small groups of X-ray stars outside of major clusters,
the results are mixed (see Tab.~\ref{groups-ir.tab}). For nine
of the 16 small groups in the HAWK-I field, small infrared clusterings
are clearly apparent in our images; for two cases the situation remains unclear,
and for five small X-ray groups we see no indication of corresponding
infrared clusterings.
The general level of agreement (i.e.~the infrared confirmation of 
$\sim 80\%$ of the reported X-ray clusterings)
appears quite good.

\section{Summary and conclusions}

Our deep HAWK-I images of the central parts of the CNC
provide crucial information about the infrared counterparts of the
X-ray sources detected in the $Chandra$ Carina Survey.
Being $\sim 5$~mag deeper than the 2MASS point source catalog, 
our HAWK-I source catalog
raises the fraction of known infrared counterparts to X-ray sources
from $\sim 45\%$ to
nearly $90\%$. The near-infrared properties of the X-ray counterparts
are consistent with the
results of the automatic X-ray source classification and  
show that the remaining contamination in the sample of 10\,714 X-ray sources
classified as pre-main sequence stars in the CNC is very low ($\lesssim 7\%$).

Our analysis of the near-infrared data shows that the
$K_s$-band luminosity function for the X-ray selected young stars
in the CNC agrees  (within the completeness limits) well 
to that of the young stars in the Orion Nebula Cluster.
This suggests that the initial mass function in the CNC
is similar to that in Orion. 
It also directly supports the estimate in \citet{Feigelson11}
(derived by  scaling of the X-ray luminosity functions)
of a total population of $\sim  104\,000$ young stars
in the CNC, which clearly shows that the CNC contains a very large
population of  low mass stars.

We find considerable variations in the near-infrared excess fractions
for the different clusters  in the CNC.
Whereas 32\% of the X-ray selected stars in the $\lesssim 1$~Myr old 
``Treasure Chest'' cluster show  near-infrared excesses,
these fraction are 10\% for the $\sim 1\!-\!2$~Myr old cluster Tr~14,
6\% for the $\sim 3$~Myr old cluster Tr~16, and 
only 2\% for the $\sim 5\!-\!10$~Myr old cluster Tr~15.
While the
temporal decay of 
excess fraction with increasing cluster age is qualitatively similar 
to what is found for other galactic clusters, 
the absolute values of the near-infrared excess fractions for the
clusters in the CNC are clearly lower than those typical for
nearby, less massive clusters of similar age \citep[e.g.,][]{Briceno07}. This
suggests that the process of circumstellar disk dispersal proceeds on
a faster timescale in the CNC than in the more quiescent regions, and
is most likely the consequence of the very high level of massive star
feedback in the CNC.

Finally, we note that the infrared data presented in this paper
are also used in the more detailed studies of individual
clusters in the CNC in the context of the 
CCCP
\cite[see][]{Wang11,Wolk11}.

\vspace{1cm}



\acknowledgments
This work is supported by \textit{Chandra} X-ray Observatory grant GO8-9131X (PI:~L.\ Townsley) 
and by the ACIS Instrument Team contract SV4-74018 (PI:~G.\ Garmire), issued by the 
{\em Chandra} X-ray Center, which is operated by the Smithsonian Astrophysical Observatory 
for and on behalf of NASA under contract NAS8-03060.
The near-infrared observations for this project were
collected with the HAWK-I
instrument on the VLT at Paranal Observatory, Chile, under
ESO program 60.A-9284(K).
We thank the ESO staff (especially Markus Kissler-Patig and
 Monika Petr-Gotzens) for performing these observations
in service mode and Eric Feigelson for numerous helpful comments
on a draft of this paper.
This work was supported by the German
\emph{Deut\-sche For\-schungs\-ge\-mein\-schaft, DFG\/} project
number PR 569/9-1. Additional support came from funds from the Munich
Cluster of Excellence: ``Origin and Structure of the Universe''.
This publication makes use of data products from the Two Micron All Sky Survey, which is a joint project of the University of Massachusetts and the Infrared Processing and Analysis Center/California Institute of Technology, funded by the National Aeronautics and Space Administration and the National Science Foundation.



{\it Facilities:} \facility{CXO (ASIS)}.

\clearpage

\clearpage

\clearpage

\clearpage

\clearpage

\clearpage

\clearpage

\clearpage

\begin{deluxetable}{rrrrrrr}
\centering \tabletypesize{\small} \tablewidth{0pt}
\tablecolumns{7}
\tablecaption{Near-infrared Magnitudes for the 7472 \textit{Chandra} X-ray Sources
in the Area of the HAWK-I mosaic.\label{magnitudes.tab}}

\tablehead{
\colhead{CCCP number} & \colhead{$J$ (mag)} & \colhead{$\sigma_J$ (mag)} &
\colhead{$H$ (mag)} & \colhead{$\sigma_H$ (mag)}&
\colhead{$K_s$ (mag)} & \colhead{$\sigma_{K_s}$ (mag)}}

\startdata
  1719&    11.26&   0.02&    10.63&   0.02&    10.47&   0.02\\
  1730&    -9.99&   0.00&    -9.99&   0.00&    -9.99&   0.00\\
  1786&    -9.99&   0.00&    19.64&   0.09&    18.56&   0.07\\
  1844&    15.78&   0.01&    15.16&   0.01&    14.97&   0.01\\
  1846&    14.51&   0.01&    13.54&   0.01&    13.20&   0.01\\
\enddata

\tablecomments{
Table 1 is available in its entirety 
in the electronic edition of the
Astrophysical Journal Supplement. 
A portion is
shown here for guidance regarding its form and content.
The listed errors are the statistical errors reported from
 photometry routine and do not include possible systematic
 uncertainties (see discussion in the text).
A value of -9.99 means that no magnitude value is available.
}
\end{deluxetable}

\begin{deluxetable}{rcl}
\centering \tabletypesize{\small} \tablewidth{0pt}
\tablecolumns{3}
\tablecaption{Infrared counterparts of the small groups of X-ray stars\label{groups-ir.tab}}

\tablehead{
\colhead{Group number} & \colhead{RA (J2000) Dec} & \colhead{IR} }

\startdata

 1& 10:42:13.5 \,  -59:35:59 &  outside HAWK-I\\
 2& 10:42:25.8 \,  -59:46:21 &  outside HAWK-I\\
 3& 10:42:46.7 \,  -59:46:56 &  outside HAWK-I\\
 4& 10:42:47.3 \,  -59:25:35 &  outside HAWK-I\\
 5& 10:42:49.5 \,  -59:09:05 &  outside HAWK-I\\
 6& 10:42:52.8 \,  -60:13:20 &  outside HAWK-I\\
 7& 10:42:53.4 \,  -59:26:15 &  no IR cluster\\
 8& 10:43:34.6 \,  -60:00:18 &  outside HAWK-I\\
 9& 10:43:41.0 \,  -60:01:16 &  outside HAWK-I\\
10& 10:43:46.6 \,  -60:02:24 &  outside HAWK-I\\
11& 10:43:59.9 \,  -60:01:59 &  outside HAWK-I\\
12& 10:44:15.0 \,  -60:00:05 &  outside HAWK-I\\
13& 10:44:20.8 \,  -59:59:03 &  outside HAWK-I\\
14& 10:44:22.5 \,  -59:59:37 &  outside HAWK-I\\
15& 10:44:22.6 \,  -59:25:14 &  no IR cluster\\
16& 10:44:26.3 \,  -59:59:55 &  outside HAWK-I\\
17& 10:44:40.2 \,  -59:46:52 &  small group \\
18& 10:44:51.9 \,  -60:25:09 &  outside HAWK-I\\
19& 10:44:56.7 \,  -59:24:51 &  no IR cluster\\
20& 10:44:58.5 \,  -59:47:11 &  small embedded group\\
21& 10:45:30.0 \,  -59:57:41 &  no IR cluster\\
22& 10:45:36.8 \,  -59:57:58 &  low-density group\\
23& 10:45:38.5 \,  -60:00:44 &  loose group\\
24& 10:45:40.4 \,  -60:01:07 &  unclear, part of group 23 ?\\
25& 10:45:44.9 \,  -59:55:27 &  loose group of bright stars\\
26& 10:46:05.8 \,  -59:56:55 &  loose group \\
27& 10:46:06.0 \,  -59:58:59 &  no IR cluster\\
28& 10:46:50.7 \,  -60:04:23 & loose clustering\\
29& 10:46:52.0 \,  -60:06:03 & loose clustering \\
28& 10:46:54.4 \,  -60:04:39 & dense clustering\\
31& 10:47:17.4 \,  -60:07:51 & unclear\\
\enddata
\end{deluxetable}

\end{document}